\documentclass[fleqn,twoside]{article}
\usepackage{espcrc2}

\usepackage{epsfig}

\newcommand{\Tr}{\mbox{Tr}}
\newcommand{\VEV}[1]{\left\langle #1\right\rangle}
\newcommand{\showpsfig}[2]{\centerline{\epsfig{file=#1,width=#2,clip=}}}
\title{Topological Susceptibility with Three Flavors of Staggered Quarks}
\author{ MILC Collaboration: C.~Aubin
\address{Department of Physics, Washington University, St.~Louis, MO 63130, USA},
C.~Bernard$\,\null^{\rm a}$,
Brian Billeter
\address{Physics Department, University of Utah, Salt Lake City, UT
  84112, USA},
C.~DeTar$\,\null^{\rm b}$\thanks{Presented by C.~DeTar.},
Steven~Gottlieb
\address{Department of Physics, Indiana University, Bloomington, IN 47405, USA},
E.B.~Gregory
\address{Department of Physics, University of Arizona, Tucson, AZ 85721, USA}, 
U.M.~Heller
\address{American Physical Society, One Research Road, Box 9000, Ridge, NY 11961, USA},
J.E.~Hetrick
\address{Physics Department, University of the Pacific, Stockton, CA 95211, USA},
J.~Osborn$\,\null^{\rm b}$,
R.L.~Sugar
\address{Department of Physics, University of California, Santa Barbara, CA 93106, USA},
and D.~Toussaint$\,\null^{\rm d}$
} %end \author

\begin{document}
%%%%%%%%%%%%%%%%%%%%%%%%%%%%%%%%%%%%%%%%%%%%%%%%%%%%%%%%%%%%%%%%%%%%%%
%   Abstract
%%%%%%%%%%%%%%%%%%%%%%%%%%%%%%%%%%%%%%%%%%%%%%%%%%%%%%%%%%%%%%%%%%%%%%
\begin{abstract}
As one test of the validity of the staggered-fermion fourth-root
determinant trick, we examine the suppression of the topological
susceptibility of the QCD vacuum in the limit of small quark mass.
The suppression is sensitive to the number of light sea quark
flavors. Our study is done in the presence of 2+1 flavors of dynamical
quarks in the improved staggered fermion formulation.
Variance-reduction techniques provide better control of statistical
errors.  New results from staggered chiral perturbation theory account
for taste-breaking effects in the low-quark mass behavior of the
susceptibility, thereby reducing scaling violations from this source.
Measurements over a range of quark masses at two lattice spacings
permit a rough continuum extrapolation to remove the remaining lattice
artifacts.  The results are consistent with chiral perturbation
theory with the correct flavor counting.
\end{abstract}

\maketitle
%%%%%%%%%%%%%%%%%%%%%%%%%%%%%%%%%%%%%%%%%%%%%%%%%%%%%%%%%%%%%%%%%%%%%%
%   Introduction
%%%%%%%%%%%%%%%%%%%%%%%%%%%%%%%%%%%%%%%%%%%%%%%%%%%%%%%%%%%%%%%%%%%%%%
\section{INTRODUCTION}

The continuum Euclidean partition function for QCD, restricted to
topological charge sector $\nu$ with $n_f$ degenerate flavors is
\begin{eqnarray}
   Z_\nu &=& \int [dU] \, \exp[S(U)] \det[M(U)]^{n_f}  \nonumber \\ 
   \det[M(U)] &=& m^{|\nu|}\prod_n(\lambda_n^2 + m^2)
\label{eq:partition}
\end{eqnarray}
where $U$ is the gauge field, $S(U)$, the Yang-Mills action, and
$\lambda_n$ are the nonzero eigenvalues of the Dirac operator
$D(U)$, where $M(U) = D(U) + m$ \cite{ref:LS}.  As the quark
mass $m$ vanishes, nonzero values of $\nu$ are suppressed and the
topological susceptibility
\begin{displaymath}
   \chi = \VEV{\nu^2}/V = \sum_{\nu} \nu^2 Z_\nu/ZV
\end{displaymath}
for Eucildean four-volume $V$ vanishes.  Chiral perturbation theory
predicts
\begin{equation}
   1/\chi = 2/\mu f^2 \sum_i n_i/m_i
\label{eq:suscept.vs.m}
\end{equation}
for several flavors of degeneracy $n_i$ and mass $m_i$.

The lattice partition function for staggered fermions has a similar
form.  But to compensate for the unwanted four-fold ``taste''
degeneracy, it is common to replace $\det[M(U)]^{n_f}$ by
$\det[M(U)]^{n_f/4}$ (fourth-root trick) with the hope that the
continuum limit recovers Eq.\ (\ref{eq:partition}).  Two principal lattice
artifacts modify Eq.\ (\ref{eq:suscept.vs.m}) \cite{ref:milc.hart}.
First, topological charge counting becomes ambiguous on the scale of
the lattice spacing.  Second, the spectrum of eigenvalues of the
fermion matrix $M(U)$ is modified so that would-be zero modes are
shifted away from zero and the four-fold taste degeneracy required by
the fourth-root trick is only approximate \cite{ref:stagg.eigenvalue}.

The staggered fermion artifacts can now be accommodated within the
framework of staggered chiral perturbation theory
\cite{ref:brief.report}, and, as we show from a numerical study
with improved staggered fermions, improves the approach to the
continuum limit, where we find a result consistent with Eq.\
(\ref{eq:suscept.vs.m}).

\begin{table}[t]
\begin{tabular}{|l|l|l|r|}
\hline
$am_{u,d}$ / $am_s$  & \hspace{-1.0mm}$10/g^2$  & $L$ & lats. \\
\hline                             
quenched       & 8.00  & 20 & 408   \\
0.05  / 0.05   & 6.85  & 20 & 425    \\
0.04  / 0.05   & 6.83  & 20 & 351    \\
0.03  / 0.05   & 6.81  & 20 & 564    \\
0.02  / 0.05   & 6.79  & 20 & 484    \\
0.01  / 0.05   & 6.76  & 20 & 658    \\   
0.007  / 0.05  & 6.76  & 20 & 493    \\   
\hline
\end{tabular}
\vspace*{-5mm}
\label{tab:coarse}
\end{table}

\section{STAGGERED CHIRAL PERTURBATION THEORY}

We summarize the derivation of the leading S$\chi$PT formula for the
susceptibility and refer the reader to Ref.~\cite{ref:brief.report} for
details.  We start from the Aubin and Bernard generalization of the
formalism of Lee and Sharpe \cite{ref:ABLS} to obtain the effective
chiral Lagrangian for staggered fermions to ${\cal O}(a^2)$:
\begin{eqnarray*}
  {\cal L} &=& 
   \frac{f^2}{8} \Tr \left(\partial_\mu U^\dagger \partial_\mu U \right)
    - \frac{\mu f^2}{4} \Tr[{\cal M}(U^\dagger + U)] \\
   &+& {m_0^2 \over 2}\phi_{0I}^2
   + a^2 {\cal V}(U)
\end{eqnarray*}
The chiral field $U$ describes $N$ flavors of 4 tastes each and
$\phi_{0I}$ is the flavor-singlet, taste-singlet field, ${\cal M}$ is
the taste-degenerate quark mass matrix, and ${\cal V}(U)$ is the
dimension-six potential \cite{ref:ABLS}, which breaks the continuum
taste symmetry.  Only one Goldstone pion in the sixteen-member
taste-multiplet survives.  The ``pion'' with the highest mass is a
taste singlet.

\begin{table}[t]
\begin{tabular}{|l|l|l|r|}
\hline
$am_{u,d}$ / $am_s$  & \hspace{-1.0mm}$10/g^2$  & $L$ & lats. \\
\hline                             
quenched        & 8.40  & 28 & 396  \\
0.031  / 0.031  & 7.18  & 28 & 496  \\   
0.0124  / 0.031 & 7.11  & 28 & 527  \\   
0.0062  / 0.031 & 7.09  & 28 & 592  \\   
0.0031  / 0.031 & 7.08  & 40 &  80  \\   
\hline
\end{tabular}
\vspace*{-5mm}
\label{tab:fine}
\end{table}

We work to quadratic order in the fields and in the mean-field
approximation and follow Leutwyler and Smilga \cite{ref:LS} to obtain
the susceptibility with a degeneracy of $n_i$ flavors of four tastes
each:
\begin{displaymath}
  \chi = \VEV{\nu^2}/V = 
     \frac{f^2/16}{\sum_{i=1} n_i (1/m^2_{iI} + 1/m_0^2)}
\end{displaymath}
where the bare flavor-neutral, taste-singlet pseudoscalar masses are
\begin{displaymath}
    m^2_{iI} = 2 \mu m_i + a^2 \Delta_I
\end{displaymath}
in terms of the bare quark masses $m_i$ and the taste splitting $\Delta_I$.
To eliminate the taste degeneracy, leaving only two degenerate $u$ and
$d$ quarks and one $s$ quark we put $n_{ud} = 1/2$ and $n_s = 1/4$,
giving, finally
\begin{equation}
  \chi = 
     \frac{f^2 m^2_{\pi,I}/8}
      {1 +  m^2_{\pi,I}/2 m^2_{ss,I} + 3 m^2_{\pi,I}/2 m_0^2}
\label{eq:suscept_tb}
\end{equation}
This leading-order formula is identical to that of continuum chiral
perturbation theory, except that it requires the taste-singlet meson
masses.  It interpolates \cite{ref:Duerr} between the
perturbative and quenched regimes and reproduces the Witten-Veneziano
formula at infinite meson mass \cite{ref:Witten.Veneziano}.

\section{SIMULATION AND ANALYSIS}

The topological charge density was measured as described in
\cite{ref:milc.hart} on an ensemble of lattices (quenched and with up,
down, and strange improved Asqtad staggered quarks) of two lattice
spacings: ``coarse'' ($a \approx 0.125$ fm) and ``fine'' ($a \approx
0.09$ fm), as listed in the upper and lower tables respectively
\cite{ref:ensembles}.  Included in this analysis is a new fine-lattice
$40^3 \times 96$ ensemble with $u$ and $d$ quark masses at 1/10 the
nominal strange quark mass.

\begin{figure}[t]
  \vspace*{-25mm}
\centerline{
  \showpsfig{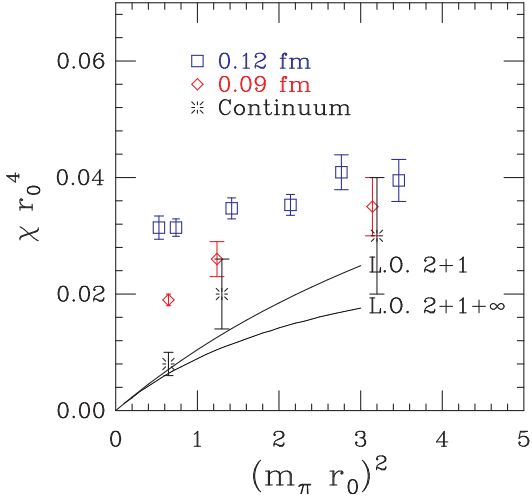}{90mm}
}
 \vspace*{-10mm}

\caption{Result of \protect\cite{ref:milc.hart}.  The curves are from
 Eq.\.~(\protect\ref{eq:suscept_tb}) with (L.O.\ $2+1+\infty$) and
 without (L.O.\ $2+1$) the $m_0^2$ term.}

\label{fig:resulta}
 \vspace*{-4mm}
\end{figure}

To determine the susceptibility, we use a variance reduction technique
\cite{ref:variance.reduction}, based on the observation that with
large lattice volumes we gain statistically by writing the
susceptibility as a local, intensive observable, rather than a global
observable:
\begin{displaymath}
 \VEV{\nu^2}/V = \sum_r C_{\rm meas}(r),
\end{displaymath}
where the measured topological charge density correlation function, is
\begin{displaymath}
  C_{\rm meas}(r) =  \VEV{\rho(0) \rho(r)}
\end{displaymath}
and $\rho$ is the topological charge density operator.  

The correlator has the asymptotic form \cite{ref:Shuryak.Verbaarschot}
\begin{displaymath}
  C_{\rm fit}(r) =  b_\eta D(m_\eta,r) + 
        b_{\eta^\prime} D(m_{\eta^\prime},r) + \ldots{}
\end{displaymath}
where $D(m,r)$ is the Euclidean scalar propagator.

We fit the correlator to this form for large $|r|$ and use it to model
the large distance behavior, thereby reducing the variance due to
contributions at large $|r|$:
\begin{displaymath}
   \VEV{\nu^2}/V = \sum_{|r| \le |r_{\rm cut}|} C_{\rm meas}(r) + 
     \sum_{|r| > |r_{\rm cut}|} C_{\rm fit}(r).
\end{displaymath}

\section{RESULTS}

\begin{figure}[t]
  \vspace*{-25mm}
\centerline{
  \showpsfig{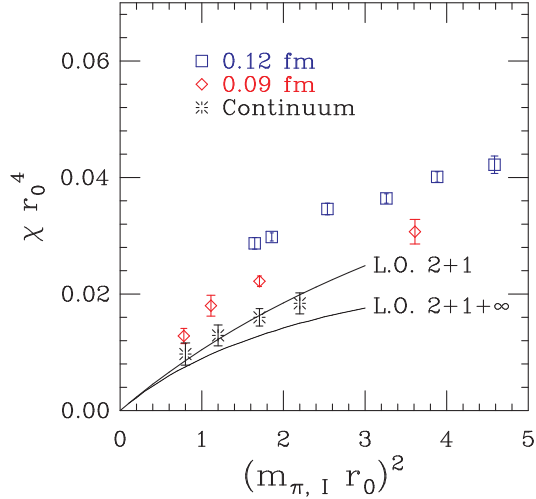}{90mm}
}
 \vspace*{-10mm}
\caption{New result.} 
\label{fig:resultb}
  \vspace*{-4mm}
\end{figure}

As we have seen, at lowest order in chiral perturbation theory the
staggered fermion artifact is removed by writing the susceptibility
as a function of the taste-singlet pion mass, rather than the
Goldstone pion mass.  The improvement can be seen by comparing
Figs.~\ref{fig:resulta} and \ref{fig:resultb}.

Residual lattice artifacts come from the lattice operator definition
of topological charge density.  It is plausible that they scale as
${\cal O}(a^2)$.  The continuum extrapolation of
Fig.~\ref{fig:resultb} at fixed abscissa is based on that assumption.

%%%%%%%%%%%%%%%%%%%%%%%%%%%%%%%%%%%%%%%%%%%%%%%%%%%%%%%%%%%%%%%%%%%%%%
%   Acknowledgments
%%%%%%%%%%%%%%%%%%%%%%%%%%%%%%%%%%%%%%%%%%%%%%%%%%%%%%%%%%%%%%%%%%%%%%

This work is supported by the US NSF and DOE. Computations were done
with Fermilab/SciDAC resources and at NCSA.  Gauge configurations
were generated at SDSC, ORNL, PSC, and NCSA.

%%%%%%%%%%%%%%%%%%%%%%%%%%%%%%%%%%%%%%%%%%%%%%%%%%%%%%%%%%%%%%%%%%%%%%
%   References
%%%%%%%%%%%%%%%%%%%%%%%%%%%%%%%%%%%%%%%%%%%%%%%%%%%%%%%%%%%%%%%%%%%%%%

%%%%%%%%%%%%%%%%%%%%%%%%%%%%%%%%%%%%%%%%%%
\end{document}